\documentclass{iau}
\usepackage{graphicx}

\title[Binary and multiple magnetic Ap/Bp stars] 
{Binary and multiple magnetic Ap/Bp stars}

\author[Denis Rastegaev et al.]   
{Denis Rastegaev$^1$,
Yuri Balega, Vladimir Dyachenko, Alexander Maksimov \and Evgenij Malogolovets}

\affiliation{Special Astrophysical Observatory, \\ Nizhnij Arkhyz,
Zelenchukskiy region, Karachai-Cherkessian Republic, Russia 369167 \\ $^1$email: {\tt leda@sao.ru} \\[\affilskip]}

\pubyear{2013}
\volume{302}  
\pagerange{119--126}
\jname{Magnetic fields throughout stellar evolution}
\editors{A.C. Editor, B.D. Editor \& C.E. Editor, eds.}
\begin{document}

\maketitle

\begin{abstract}
We present the results of speckle interferometric observations of 273 magnetic stars most of which are Ap/Bp type. All observations were made at the 6-m telescope of the Special Astrophysical Observatory of the Russian Academy of Sciences. We resolved 58 binary and 5 triple stars into individual components. Almost half of these stars were astrometrically resolved for the first time. The fraction of speckle interferometric binaries/multiples in the sample of stars with confirmed magnetic fields is 23\%. We expect that the total fraction of binaries/multiples in the sample with account for spectroscopic short-period systems and wide common proper motion pairs can be twice higher. The detected speckle components have a prominent peak in the $\rho$ distribution that corresponds to the closest resolved pairs.                                                                                                         
\end{abstract}

\firstsection 
\section{Introduction}
Chemically peculiar stars of spectral classes A and B are objects in the atmospheres of which there are anomalies of chemical elements such as Sr, Cr, Eu, Si, etc. These stars often have global magnetic fields which are considered to be responsible for the chemical anomalies. The existing empirical material does not allow us to make the final choice in favor of one of the theories regarding the origin and evolution of magnetic fields. Usually, studies of stellar magnetism use spectroscopic and photometric observations. The accumulated wealth of observational data however does not answer the main question about the formation of a global magnetic field of the star. We propose to use another source of information --- an interferometric study of binary and multiple magnetic stars. According to current views most stars are born in groups of two or more components. Single stars are formed by the decay of such groups. Compared with single systems, binary and multiple systems have three additional quantities: the orbital angular momentum, the eccentricity of the orbit and the mass ratio of the components. These values give us important information about the star formation process and physical conditions in protostellar matter including the presence of a magnetic field. The study of binary and multiple magnetic stars can be a cornerstone in the issue of the origin of stellar magnetic fields. To address this question we conducted speckle interferometric observations with high angular resolution of all known magnetic stars with global fields in the Northern Hemisphere (273 objects with $\delta>-30^{\circ}$). Observations were made at the 6 m telescope BTA, the largest optical telescope in Eurasia.

\section{Sample}

A sample of objects for observations was based on the Catalog of Magnetic Stars (Romanyuk \& Kudryavtsev 2008). It contains a list of 355 chemically peculiar objects (mostly Ap/Bp) with detected global magnetic fields. We added 17 new magnetic stars discovered after the publication of the catalog. Therefore the total number of stars in the sample is 372. For the majority of stars in the list (322 objects) only the value of the longitudinal component of the field Be is known. For 48 stars the surface fields are determined from the splitting of Zeeman components. The vast majority of the sample objects are brighter than 10$^m$ in the $V$-band. The stars are uniformly distributed on the celestial sphere, although a relatively small number (about 20\%) of objects belong to open clusters of different ages. The BTA can capture only 273 objects from our sample with declinations $\delta>-30^{\circ}$.

\begin{figure}
\begin{center}
 \includegraphics[width=5.4in]{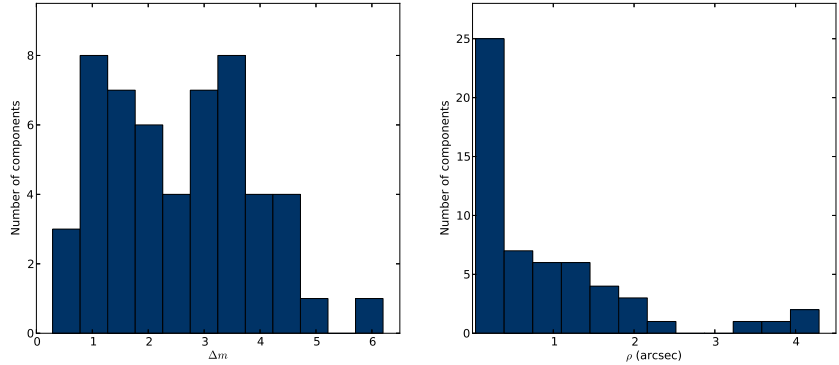} 
 \caption{Magnitude difference and angular separation distributions for resolved pairs.}
   \label{fig1}
\end{center}
\end{figure}

\section{Observations}
The speckle interferometric observations of 273 magnetic CP stars were carried out at the BTA in 2009--2012. They were performed with the speckle interferometer engineered at the SAO RAS (Maksimov et al. 2009). We used the PhotonMAX512 camera based on an internal electron multiplying CCD97 (EMCCD) produced by Princeton Instruments with a $512\times512$ pixel array. The limiting magnitude of our speckle interferometer is $\approx15^{m}$ in the $V$-band depending on seeing conditions. Basically we employed two filters: 550/20 and 800/100 nm (central wavelength/bandwidth). We took 2000 short exposure images in each filter for almost all observed objects. High quantum efficiency and linearity of the detector permits the maximum magnitude difference between the components to reach up to 5-6$^m$ depending on angular separation and weather conditions (Fig. 1). The minimum angular separation between the components is determined by the diffraction limit of the 6 m telescope. It is 0.022$''$ and 0.033$''$ for 550/20 and 800/100 nm filters, respectively. The size of the detector's field $4.4\times4.4''$ allowed secondary components to be discovered at angular separations as large as 3$''$ from the primary star. The accuracy of our speckle interferogram processing method may be as good as 0.02$^{m}$, 0.001$''$, and 0.1$^{\circ}$ for the component magnitude difference, angular separation, and position angle, respectively.

\section{Results}

For 63 stars in our sample, we observed speckle interferometric companions. Among the resolved systems 58 are binaries and 5 are triples. Twenty nine companions were resolved astrometrically for the first time. The fraction of speckle interferometric binaries/multiples in the sample of 273 stars with confirmed magnetic fields is 23\%. Magnitude difference and angular separation distributions for resolved pairs are shown in Fig. 1. To plot these histograms we used 56 measurements of $\rho$ and 53 $\Delta$m. We want to draw attention to the unusual profile of the $\rho$ distribution. Speckle interferometric components of magnetic stars tend to be located close to the primary star. Half of the resolved stars have companions with $\rho<0.32''$. This result is not a selection effect because close interferometric components are harder to detect than the wide pairs. The distribution of $\Delta$m for 28 resolved speckle companions with $\rho<0.32''$ resembles that on the left half of Fig. 1. Reconstructed images of six systems resolved for the first time on BTA are presented in Fig. 2. The table below is a list of all the stars that have speckle components. The systems resolved astrometrically for the first time are marked in bold.

\begin{figure}
\begin{center}
 \includegraphics[width=5.4in]{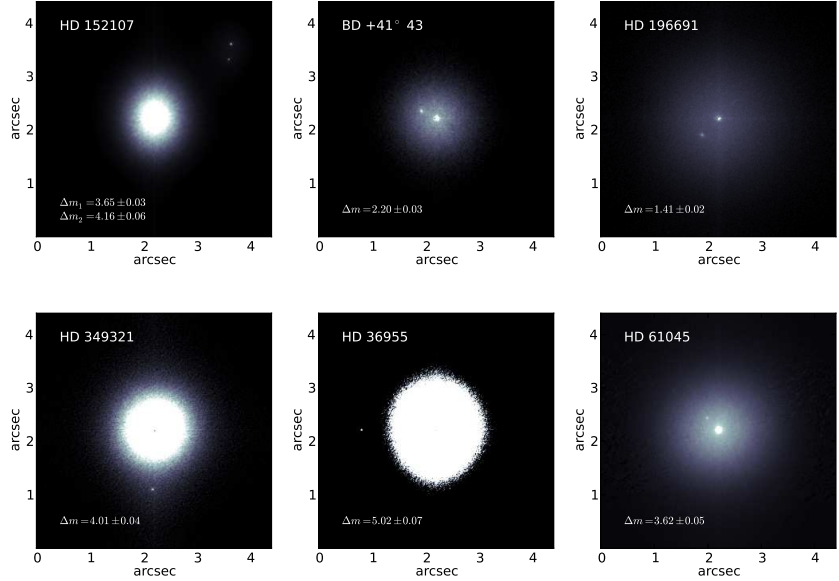} 
 \caption{Bispectrum image reconstruction of 1 triple and 5 binary systems. All presented binaries are resolved for the first time. The magnitude difference between components is marked for each system. The images are obtained in the 800/100 nm filter.}
   \label{fig2}
\end{center}
\end{figure}

\begin{table}
  \begin{center}
  \caption{Resolved stars.}
  \label{tab1}
 {\scriptsize
  \begin{tabular}{|l|l|l|l|l|l|l|l|}\hline 
{\bf HD 965}&{\bf BD+41 43}&{\bf HD 2887}&BD+40 175&{\bf HD 5797}&HD 6757&{\bf HD 8855}&{\bf HD 10783} \\ \hline
HD 12447&HD 15089&{\bf HD 16605}&HD 16728&{\bf HD 21699}&HD 22470&HD 29009&{\bf HD 30466} \\ \hline
{\bf HD 293764}&HD 35100&HD 35456&{\bf HD 35502}&HD 36313&{\bf HD 36540}&{\bf HD 36955}&HD 37017 \\ \hline
HD 37140&{\bf HD 37479}&HD 40312&HD 258686&{\bf HD 51418}&{\bf HD 61045}&HD 65339&{\bf HD 64486} \\ \hline
HD 78316&{\bf HD 79158}&HD 81009&{\bf HD 89069}&HD 98088&HD 99563&{\bf HD 103498}&{\bf HD 108651} \\ \hline
HD 130559&HD 137909&{\bf HD 144334}&HD 145501&HD 148112&HD 152107&HD 158450&{\bf HD 164258} \\ \hline
HD 170000&HD 169887&{\bf HD 349321}&{\bf HD 343872}&{\bf HD 338226}&BD+35 3616&{\bf HD 184471}&HD 192913 \\ \hline
HD 196178&{\bf HD 196691}&HD 200177&HD 201601&HD 210432&{\bf HD 213918}&HD 217833& \\ \hline
  \end{tabular}
  }
 \end{center}
\end{table}

\section{Conclusion}
According to our research the fraction of speckle interferometric binary and multiple systems in the sample of 273 CP stars with confirmed magnetic fields makes up 23\% without account for undetected companions. Generally the speckle interferometric components have orbital periods larger than spectroscopic and smaller than common proper motion pairs. We expect that the total fraction of binaries/multiples in the sample with account for spectroscopic short-period systems and wide common proper motion pairs can be twice higher. The detected speckle components have a prominent peak in the $\rho$ distribution that corresponds to the closest resolved pairs. More detailed and refined results of the presented study will be published soon.

{\scriptsize
\acknowledgements
This work was supported by Federal Target Program "Scientific and scientific-pedagogical personnel of innovative Russia" for 2009--2013 years (N 8704) and grant of the President of the Russian Federation for the state support of young Russian PhD scientists (MK-1001.2012.2).
}

\end{document}